\documentclass[reprint, amsfonts, amsmath, showpacs,aps,superscriptaddress]{revtex4-1}
\usepackage{bm}
\usepackage{graphicx}
\usepackage{epstopdf}
\usepackage{color}

\begin{document}
\title{Evidence for a dynamical ground state in a frustrated pyrohafnate \bm{${\rm Tb_2Hf_2O_7}$}}
\author{V. K. Anand}
\altaffiliation{vivekkranand@gmail.com}
\affiliation{\mbox{Helmholtz-Zentrum Berlin f\"{u}r Materialien und Energie GmbH, Hahn-Meitner Platz 1, D-14109 Berlin, Germany}}
\author{L. Opherden}
\affiliation{\mbox{Dresden High Magnetic Field Laboratory (HLD-EMFL), Helmholtz-Zentrum Dresden-Rossendorf,} 01328 Dresden, Germany}
\affiliation{Institut f\"ur Festk\"orper- und Materialphysik, TU Dresden, 01062 Dresden, Germany}
\author{J. Xu}
\affiliation{\mbox{Helmholtz-Zentrum Berlin f\"{u}r Materialien und Energie GmbH, Hahn-Meitner Platz 1, D-14109 Berlin, Germany}}
\author{D. T. Adroja}
\affiliation{ISIS Facility, Rutherford Appleton Laboratory, Chilton, Didcot, Oxon, OX11 0QX, United Kingdom}
\affiliation{\mbox{Highly Correlated Matter Research Group, Physics Department, University of Johannesburg, P.O. Box 524,}
Auckland Park 2006, South Africa}
\author{A. D. Hillier}
\author{P. K. Biswas}
\affiliation{ISIS Facility, Rutherford Appleton Laboratory, Chilton, Didcot, Oxon, OX11 0QX, United Kingdom}
\author{T.~Herrmannsd\"orfer}
\author{M. Uhlarz}
\affiliation{\mbox{Dresden High Magnetic Field Laboratory (HLD-EMFL), Helmholtz-Zentrum Dresden-Rossendorf,} 01328 Dresden, Germany}
\author{J. Hornung}
\author{J. Wosnitza}
\affiliation{\mbox{Dresden High Magnetic Field Laboratory (HLD-EMFL), Helmholtz-Zentrum Dresden-Rossendorf,} 01328 Dresden, Germany}
\affiliation{Institut f\"ur Festk\"orper- und Materialphysik, TU Dresden, 01062 Dresden, Germany}
\author{E. Can\'evet}
\affiliation{\mbox{Laboratory for Neutron Scattering and Imaging, Paul Scherrer Institute, CH-5232 Villigen PSI, Switzerland}}
\affiliation{Department of Physics, Technical University of Denmark, 2800 Kgs. Lyngby, Denmark}
\author{B. Lake}
\altaffiliation{bella.lake@helmholtz-berlin.de}
\affiliation{\mbox{Helmholtz-Zentrum Berlin f\"{u}r Materialien und Energie GmbH, Hahn-Meitner Platz 1, D-14109 Berlin, Germany}}

\date{\today}

\begin{abstract}
We report the physical properties of Tb$_2$Hf$_2$O$_7$ based on ac magnetic susceptibility $\chi_{\rm ac}(T)$, dc magnetic susceptibility  $\chi(T)$, isothermal magnetization $M(H)$, and heat capacity $C_{\rm p}(T)$ measurements combined with muon spin relaxation ($\mu$SR) and neutron powder diffraction measurements. No evidence for long-range magnetic order is found down to 0.1~K\@. However, $\chi_{\rm ac}(T)$ data present a frequency-dependent broad peak (near 0.9~K at 16~Hz) indicating slow spin dynamics. The slow spin dynamics is further evidenced from the $\mu$SR data (characterized by a stretched exponential behavior) which show persistent spin fluctuations down to 0.3~K\@. The neutron powder diffraction data collected at 0.1~K show a broad peak of magnetic origin (diffuse scattering) but no magnetic Bragg peaks. The analysis of the diffuse scattering data reveals a dominant antiferromagnetic interaction in agreement with the negative Weiss temperature. The absence of long-range magnetic order and the presence of slow spin dynamics and persistent spin fluctuations together reflect a dynamical ground state in Tb$_2$Hf$_2$O$_7$.
\end{abstract}

\maketitle

\section{\label{Intro} INTRODUCTION}

The discovery of spin-ice behavior in Dy$_2$Ti$_2$O$_7$ and Ho$_2$Ti$_2$O$_7$ and associated magnetic monopole dynamics have created a great research interests in the rare-earth pyrochlore oxides $R_2B_2$O$_7$ ($R$ is a trivalent rare earth ion and $B$ a tetravalent transition-metal ion or Ge, Sn, Pb) \cite{Harris1997, Ramirez1999,  Siddharthan1999, Hertog2000, Bramwell2001,Castelnovo2008, Morris2009, Gardner2010, Castelnovo2012, Gingras2014}. These materials contain corner-sharing tetrahedra of $R^{3+}$ ions which provide a very suitable condition for realizing frustrated magnetism. With this atomic arrangement, the interplay of crystal electric field (CEF), antiferromagnetic exchange, and ferromagnetic dipolar interactions lead to diverse magnetic states in these materials \cite{Gardner2010}. The spin-ice state is realized with the frustrated `two-in/two-out' spin configuration resulting from ferromagnetic interactions with local $\langle 111 \rangle $ Ising-anisotropy \cite{Harris1997}. On the other hand, antiferromagnetic interactions result in an `all-in/all-out' spin configuration  \cite{Melko2004,Anand2015, Lhotel2015, Xu2015, Bertin2015}.  Spin-liquid behavior is another interesting magnetic ground state that has been observed in Tb$_2$Ti$_2$O$_7$ \cite{Gardner2010, Gingras2014}. 

Tb$_2$Ti$_2$O$_7$, like the spin-ice compounds Dy$_2$Ti$_2$O$_7$ and Ho$_2$Ti$_2$O$_7$, also has a CEF-split doublet ground state and displays $\langle 111 \rangle $ Ising anisotropy. Despite a large negative Weiss temperature ($\theta_{\rm p} = -19$~K) no signature of long-range magnetic order has been found in Tb$_2$Ti$_2$O$_7$ down to 50 mK \cite{Gardner1999,Gardner2001,Gardner2003,Gingras2000}. As the effective interaction in Tb$_2$Ti$_2$O$_7$ is antiferromagnetic in nature, the $\langle 111 \rangle $ anisotropy should not cause magnetic frustration, accordingly a long-range ordered magnetic ground state is expected. In order to explain the low temperature properties of Tb$_2$Ti$_2$O$_7$ two theoretical scenarios have been proposed: a) a quantum spin ice (QSI) scenario \cite{Molavian2007,Molavian2012}, and b) a nonmagnetic singlet ground state scenario \cite{Bonville2011}. In the QSI scenario, virtual quantum excitations between the CEF ground state doublet and the first excited doublet ($\sim 15$~K) have been proposed to renormalize the effective low-energy spin Hamiltonian such that the ground state of Tb$_2$Ti$_2$O$_7$ is a quantum spin-ice state \cite{Molavian2007,Molavian2012}. The second scenario argues that Tb$_2$Ti$_2$O$_7$ undergoes a Jahn-Teller transition leading to a singlet ground state with a sufficiently large splitting between ground state and first excited state which prevents long-range order \cite{Bonville2011}. Recent neutron scattering investigations on Tb$_2$Ti$_2$O$_7$ have been reported to show evidence for pinch-points and Coulombic correlations which favor the QSI picture \cite{Petit2012,Fennell2012, Guitteny2013, Fritsch2013, Fritsch2014}. 

The intriguing magnetic ground state of Tb$_2$Ti$_2$O$_7$ motivated us to investigate the Hf-analog of this compound, namely, Tb$_2$Hf$_2$O$_7$. Recently, we investigated the physical properties of the hafnate pyrochlores Nd$_2$Hf$_2$O$_7$  \cite{Anand2015,Anand2017} and Pr$_2$Hf$_2$O$_7$ \cite{Anand2016}. Nd$_2$Hf$_2$O$_7$ was found to exhibit a long-range antiferromagnetic order below $T_{\rm N }\approx 0.55$~K\@ for which our neutron diffraction study revealed an all-in/all-out arrangement of Nd$^{3+}$ moments \cite{Anand2015}. Inelastic neutron scattering confirmed the Ising anisotropy of Nd$_2$Hf$_2$O$_7$ with a Kramers doublet ground state of dipolar-octupolar character \cite{Anand2017}. Interestingly, muon spin relaxation ($\mu$SR) revealed the presence of persistent dynamic spin fluctuations deep inside the ordered state because of which the magnetic moment in the ordered state  [0.62(1)~$\mu_{\rm B}$/Nd at 0.1~K] is strongly reduced \cite{Anand2017,Anand2015}. On the other hand, Pr$_2$Hf$_2$O$_7$ with a non-Kramers doublet ground state is found not to show long-range magnetic order down to 90~mK, though slow spin dynamics and spin-freezing are inferred from ac magnetic susceptibility data \cite{Anand2016}. The magnetic ground state of Pr$_2$Hf$_2$O$_7$ possesses the ingredients for quantum spin ice behavior \cite{Anand2016,Sibille2016}.

Extending our work on hafnate pyrochlores, here we present physical properties of Tb$_2$Hf$_2$O$_7$. The magnetic ion Tb$^{3+}$ ($4f^8$, $S = 3$, $L = 3$, $J = 6$) is a non-Kramers ion. The ($2J+1 = 13$)-fold degenerate free-ion ground-state multiplet $^7F_6$ of Tb$^{3+}$, when subject to crystal field in the cubic pyrochlore structure, splits into four doublets and five singlets. Structural characterizations on Tb$_2$Hf$_2$O$_7$ report the presence of disorder (defect fluorite) in the pyrochlore structure of this compound \cite{Karthik2012}. The presence of structural disorder can perturb the crystal field environment which in turn can modify the CEF states (splitting energies and levels scheme) which have a strong bearing on the magnetic ground state of such systems. The magnetic ground states of the pyrochlores Tb$_2$Ti$_2$O$_7$ and Tb$_2$Sn$_2$O$_7$ differ significantly likely on account of difference in their CEF states \cite{Zhang2014}. 

Our $\mu$SR study on Tb$_2$Hf$_2$O$_7$ shows the absence of long-range magnetic order down to 0.3~K, however persistent spin fluctuations and slow spin dynamics are inferred from the $\mu$SR data. The ac magnetic susceptibility shows a frequency dependent broad peak which also reflects a slow spin dynamics in Tb$_2$Hf$_2$O$_7$. Consistent with $\mu$SR and ac susceptibility data, the neutron powder diffraction data also suggest the absence of long-range magnetic order. The absence of magnetic ordering and presence of persistent spin fluctuations suggest a dynamical ground state in Tb$_2$Hf$_2$O$_7$. A recent work by Sibille {\it et al}.\ \cite{Sibille2017} reports a Coulomb spin-liquid behavior in this compound.

\section{\label{ExpDetails} EXPERIMENTAL DETAILS}

Polycrystalline Tb$_2$Hf$_2$O$_7$ was prepared at the Core Lab for Quantum Materials, Helmholtz-Zentrum Berlin (HZB) by solid-state reaction similar to the synthesis of pyrohafnates Nd$_2$Hf$_2$O$_7$ \cite{Anand2015} and Pr$_2$Hf$_2$O$_7$ \cite{Anand2016}. High-purity Tb$_2$O$_3$ (99.99\%) and HfO$_2$ (99.95\%), taken in stoichiometric ratio, were mixed and ground properly and fired at 1300~$^\circ$C for 50~h followed by three additional successive grindings, pelletizings, and firings at 1400~$^\circ$C, 1500~$^\circ$C, and 1550~$^\circ$C for 80~h each. The sample quality and crystallographic parameters were checked by room-temperature x-ray  powder diffraction (XRD). The XRD data were refined using the software FullProf. \cite{Rodriguez1993} 

The dynamic (ac) susceptibility was measured using a non-compensated coil connected to an LR700 (16 Hz) or a lock-in amplifier (333 Hz - 44.4 kHz) at Helmholtz-Zentrum Dresden-Rossendorf. The sample was precooled to 2 K using a PPMS and further cooled by adiabatic demagnetization of a paramagnetic salt.

The dc susceptibility and isothermal magnetization measurements in the temperature range 2 to 300~K and with magnetic fields up to 14~T were made using a vibrating sample magnetometer (VSM) option of a Quantum Design physical properties measurement system (PPMS) at the Core Lab for Quantum Materials, HZB. The heat capacity measurements between 1.8 and 300~K were performed by the adiabatic-relaxation technique using the PPMS at the Core Lab for Quantum Materials, HZB. The resistance measurement using a multimeter and resistivity option of PPMS indicated an insulating ground state for Tb$_2$Hf$_2$O$_7$. 

The neutron  powder  diffraction (ND) measurements were performed on the cold neutron diffractometer DMC at Paul Scherrer Institute (PSI), Switzerland. The powder sample was mounted using a thin-walled copper can of diameter 10 mm. The sample mount was cooled in a top-loading dilution refrigerator achieving temperatures down to 0.1~K\@. The ND data were collected at 0.1, 1.5, and 160~K using an incident neutron beam of wavelength $\lambda =2.4586$~\AA, counting for about 8~h at each temperature. 

The muon spin relaxation ($\mu$SR) measurements were performed at the ISIS facility, Rutherford Appleton Laboratory, Didcot, U.K.\ using the spectrometer MuSR. The powdered Tb$_2$Hf$_2$O$_7$ sample was mounted on a high-purity silver plate using diluted GE varnish which was then covered with a thin silver foil. A dilution refrigerator was used to cool the sample down to 0.3~K. The $\mu$SR data were collected at several temperatures between 0.3 and 20~K in zero field, and at 0.3~K, 1~K, 2~K, and 4~K in longitudinally applied fields up to 0.3~T\@. 

\section{\label{Structure} Crystallography}

\begin{figure}
\includegraphics[width=\columnwidth, keepaspectratio]{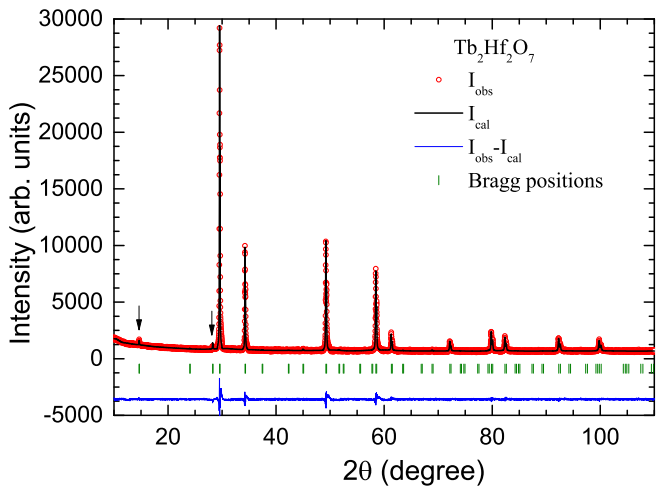}
\caption {X-ray powder diffraction pattern of polycrystalline Tb$_2$Hf$_2$O$_7$ recorded at room temperature. The solid line through the experimental points is the Rietveld refinement profile calculated for the ${\rm Eu_2Zr_2O_7}$-type face-centered cubic (space group $Fd\bar{3}m$) pyrochlore structure. The short vertical bars mark the Bragg-peak positions and the lowermost curve represents the difference between the experimental and calculated intensities. The arrows mark pyrochlore superlattice reflections.}
\label{fig:XRD}
\end{figure}

The room-temperature powder XRD pattern of Tb$_2$Hf$_2$O$_7$ is shown in Fig.~\ref{fig:XRD} along with the Rietveld refinement profile using the ${\rm Eu_2Zr_2O_7}$-type face-centered cubic  (space group $Fd\bar{3}m$) pyrochlore structure. The refinement reflects the single-phase nature of the sample without any detectable impurity. The lattice parameter is found to be $a = 10.4635(1)$~\AA\@ and the relative $x$-coordinate of O1 is $x_{\rm O1} = 0.353(2)$. The atomic position of Tb, Hf, O1, and O2 associated with the Wyckoff positions are 16d (1/2,1/2,1/2), 16c (0,0,0), 48f ($x_{\rm O1}$,1/8,1/8), and 8b (3/8,3/8,3/8), respectively.  The values of $a$ and $x_{\rm O1}$ obtained from the refinement agree very well with those reported in literature \cite{Karthik2012}. 

Even though we refined the XRD data within the pyrochlore structure, we would like to point out that Tb$_2$Hf$_2$O$_7$ does not have a well-ordered pyrochlore structure. Considering the ratio of the cation radii $r_{\rm Tb}/r_{\rm Hf} \approx 1.46 $ which is at the boundary of the values (1.46--1.78) suggested for a stable ordered pyrochlore phase  \cite{Subramanian1983},  the pyrochlore structure might not be stable and a defect fluorite phase may form as the smaller values of ratio of the cation radii have been found to favor the defect fluorite structure formation. This structural instability was noticed by Karthik {\it et al}.\ \cite{Karthik2012} through high-resolution transmission electron microscopy imaging studies. In their structural studies of $R_2$Hf$_2$O$_7$, they noticed a systematic transformation from the well-ordered pyrochlore ($Fd\bar{3}m$ space group) phase to fluorite ($Fm\bar{3}m$ space group) phase as the series progresses from La to Lu. In the  defect  fluorite  structure, the local O surroundings for both cations ($R^{3+}$ and Hf$^{4+}$, occupying cubic sites) are identical with the 8 O atoms located at tetrahedral sites  \cite{Subramanian1983}. On the other hand in a pyrochlore structure, the O surroundings of eight-fold coordinated $R^{3+}$ changes to scalenohedra (distorted cubic) consisting of 2 shorter and 6 longer $R$--O distances, whereas the O surroundings for six-fold cordinated Hf$^{4+}$ changes to trigonal antiprisms (distorted octahedral)  \cite{Subramanian1983}. Tb$_2$Hf$_2$O$_7$ lies at the boundary of pyrochlore and defect-fluorite structure. Nevertheless the characteristic pyrochlore peaks, though weak, are clearly seen in the XRD pattern (marked with arrows in Fig.~\ref{fig:XRD}). 

Very recently Sibille {\it et al}.\ \cite{Sibille2017} found evidence for Frenkel pair defects/anion disorder using combined resonant x-ray and neutron powder diffraction studies. They report $8 \pm 0.5$\% oxygen vacancy at the 48f site compensated by an oxygen occupancy of $49 \pm 3$\% at the 8a (1/8, 1/8, 1/8) site (this site is unoccupied in an ideal pyrochlore structure) \cite{Sibille2017}. In order to get a rough estimate we also tried to refine the occupancy at the 48f and 8a sites which yielded an O vacancy of $10 \pm 3$\% at the 48f site and an O occupancy of $51 \pm 9$\% for the 8a site. These numbers agree with those of Sibille et al. \cite{Sibille2017} and suggest for the presence of a similar level of disorder in both samples.

The order-disorder transition can change the crystal field environment and introduce a bond disorder, and hence influence the physical properties of a system. The crystal field anisotropy has a key role in the development of magnetic frustration on a corner sharing tetrahedra of a pyrochlore. In the case of Tb$_2$Ti$_2$O$_7$ it is the low-lying excited crystal field level which is believed to renormalize the effective Hamiltonian for quantum spin-ice state \cite{Molavian2007,Molavian2012}. The inelastic neutron scattering experiments have revealed the splitting energies between the ground state doublet and the first excited doublet in Tb$_2$Ti$_2$O$_7$ and Tb$_2$Sn$_2$O$_7$ to be 1.41~meV and 1.28~meV, respectively \cite{Zhang2014}. In contrast to discrete CEF excitations in Tb$_2$Ti$_2$O$_7$ and Tb$_2$Sn$_2$O$_7$, a rather broad CEF excitation is found in inelastic neutron scattering of Tb$_2$Hf$_2$O$_7$ for which Sibille {\it et al}.\ \cite{Sibille2017} suggest a splitting energy of at least 4.3 meV. Our heat capacity data (discussed latter) suggests a first excited state near 11 meV. This reflects a strong modification in CEF states brought by the presence of disorder in the pyrochlore structure of Tb$_2$Hf$_2$O$_7$.

\begin{figure*}
\includegraphics[width=\textwidth, keepaspectratio]{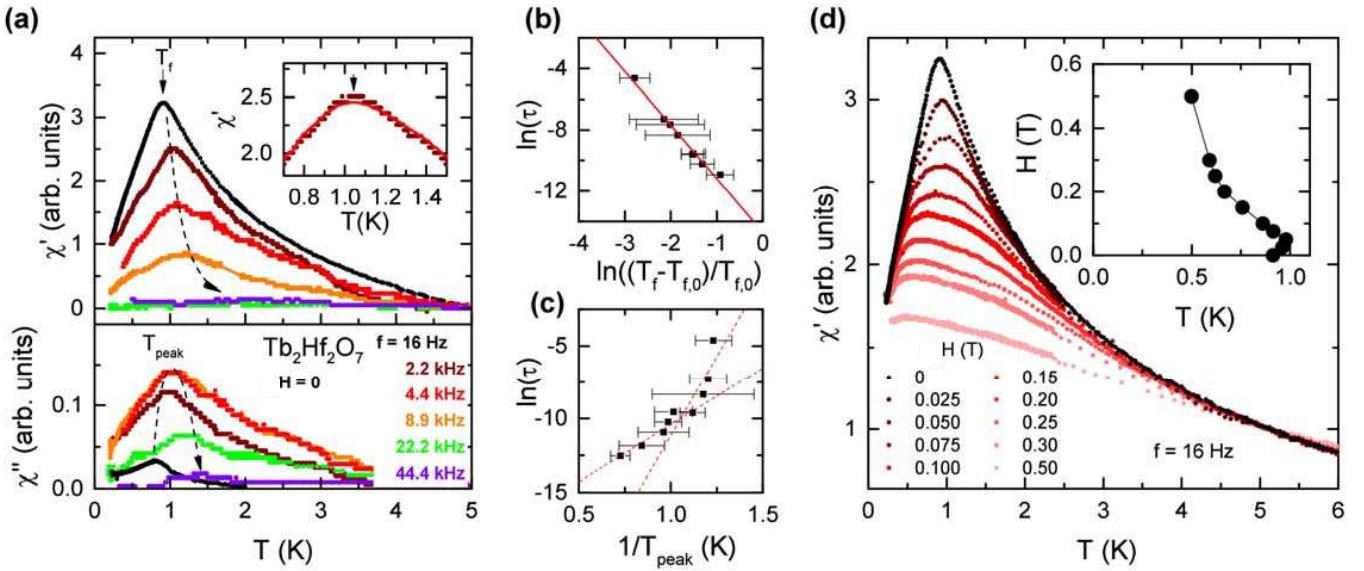}
\caption{(a) Real $\chi^\prime$ and imaginary $\chi^{\prime\prime}$ parts of the ac magnetic susceptibility of polycrystalline Tb$_2$Hf$_2$O$_7$ as a function of temperature $T$ between 0.1 and 5~K measured at the indicated frequencies $f$ and zero dc magnetic field. Inset: ac susceptibility measured at 2.2 KHz using adiabatic demagnetization cooling (dots) compared to a measurement at 2.3 kHz using zero field cooling (solid line). (b) The frequency dependence of the freezing temperature $T_f$ plotted as $\ln(\tau)$ versus $\ln(t)$, where $\tau = 1/(2\pi f)$ and the reduced temperature $t = (T_f - T_{f,0})/T_{f,0}$. The solid line represents the fit mentioned in the text. (c) Logarithmic spin-relaxation time $\tau$ versus inverse peak temperature. Dashed lines indicate Arrhenius fits. (d) $\chi^\prime(T)$ for different applied dc fields $H$ at 16 Hz under a field-cooled condition. Inset: $H$-$T$ phase-diagram for $T_f(H)$. }
\label{fig:ac_chi}
\end{figure*}

\section{\label{Sec:Chiac} \lowercase{ac} Magnetic Susceptibility}

The ac magnetic susceptibility $\chi_{\rm ac}(T)$ data of Tb$_2$Hf$_2$O$_7$ are shown in Fig.~\ref{fig:ac_chi}. Both real $\chi^\prime$ and imaginary $\chi^{\prime\prime}$ parts of $\chi_{\rm ac}(T)$ show frequency-dependent broad peaks. The peak position does not depend on the cooling protocol [inset of Fig.~\ref{fig:ac_chi}(a)]. At similar frequencies the maximum in $\chi^\prime$ and $\chi^{\prime\prime}$ coincide for cooling through adiabatic demagnetization as well as cooling under zero magnetic field using a dilution refrigerator equipped with a compensated coil-pair susceptometer. At 16 Hz, the peak in $\chi^\prime$ is centered at $T_f=0.91$~K accompanied by a peak at $T_{\rm peak}=0.81$~K in $\chi^{\prime\prime}$ [Fig.~\ref{fig:ac_chi}(a)]. With increasing frequency the peak positions of both $\chi^\prime$ and $\chi^{\prime\prime}$ shift to higher temperatures. Such frequency-dependent shift is a well-known feature of spin-glass and spin-ice systems \cite{Gardner2010,Mydosh1993}.   

The relative shift in freezing temperature per decade of frequency $\delta T_f = \Delta T_f / (T_f \Delta(\log f))= 0.06(1)$ for Tb$_2$Hf$_2$O$_7$ is comparable to that of insulating spin-glasses such as Eu$_x$Sr$_{1-x}$S and CoO$\cdot$Al$_2$O$_3$$\cdot$SiO$_2$ \cite{Mydosh1993} and the metallic ferromagnetic cluster spin-glass PrRhSn$_3$ \cite{Anand2012}. A similar value of $ \delta T_f$ (0.06--0.08) was found for Tb$_2$Ti$_2$O$_7$ \cite{Lhotel2012}. $T_f(f)$ follows a conventional power-law divergence of critical slowing down,  $\tau = \tau_0 t^{-z\nu}$ (where $\tau = 1/(2\pi f)$ and $t = (T_f - T_{f,0})/T_{f,0}$ with $T_{f,0} \approx 0.86$~K; $\nu$ being the critical exponent of the correlation length $\xi = (T_f /T_{f,0} - 1)^{-\nu}$ and $\tau\sim\xi^z$) with a critical exponent $z\nu = 3.5(2)$. A plot of $\ln(\tau)$ versus $t$ is shown in Fig.~\ref{fig:ac_chi}(b) along with the fit for parameters $z\nu$~=~3.5 and $\tau_0 = 4 \times 10^{-7}$s. Thus the ac susceptibility suggest a spin-glass type freezing in Tb$_2$Hf$_2$O$_7$. 

In order to understand the spin-relaxation in Tb$_2$Hf$_2$O$_7$,  we analyze the frequency-dependence of $T_{\rm peak}$ in the dissipative part $\chi^{\prime\prime}$. The $\tau(T)$ is shown in Fig.~\ref{fig:ac_chi}(c), plotted as $\ln(\tau)$ versus $1/T_{\rm peak}$. We find that the frequency dependence of $T_{\rm peak}$ does not follow a simple Arrhenius law, $\tau = \tau_0 \exp(E_b/k_{\rm B}T)$, where $E_b$ is the thermal energy barrier [see Fig.~\ref{fig:ac_chi}(c)]. An analysis of the Pearson correlation coefficient showed that a single Arrhenius fit leads to only 0.907 while the dynamical scaling law  leads to an excellent correlation of -0.991. Sibille {\it et al}.\ \cite{Sibille2017} also report a non-Arrhenius behavior for the spin freezing in Tb$_2$Hf$_2$O$_7$. This kind of non-Arrhenius behavior has also been observed in the spin-ice system Dy$_2$Ti$_2$O$_7$ \cite{Snyder2004} and spin-liquid system Tb$_2$Ti$_2$O$_7$ \cite{Lhotel2012} suggesting that the spin relaxation is not simply thermally activated by one energy barrier. 

We performed partial linear fits of $\ln(\tau)$ versus $1/T_{\rm peak}$ in two different $T$ regions [as shown in Fig.~\ref{fig:ac_chi}(c)] which yield the energy barriers 7.8~K and 22~K\@. For the spin-ice system Ho$_2$Ti$_2$O$_7$, an energy barrier of 27.5~K has been found by analyzing the $f$ dependence of $T_f$ by  Arrhenius law \cite{Matsuhira2000}.  For Dy$_2$Ti$_2$O$_7$ an energy barrier of 6.6~K was found to fit the spin-relaxation in the low-$T$ regime \cite{Jaubert2011}. In the case of Tb$_2$Ti$_2$O$_7$, the $f$ dependence of $T_f$  was analyzed by $\tau = \tau_0 \exp[(E_b/k_{\rm B}T)^\sigma]$ which yielded $E_b \approx 0.91$~K with $\sigma \approx 2$  \cite{Lhotel2012}.

In order to gain more insight into the spin dynamics of Tb$_2$Hf$_2$O$_7$  we also measured the ac susceptibility under applied dc magnetic fields, which is shown in Fig.~\ref{fig:ac_chi}(d). At $H \leq$ 0.05 T, $T_f$ shifts to slightly higher temperatures. However, at fields higher than 0.05~T, $T_f$  shifts to lower temperatures. The field dependence of $T_f$ is summarized in the $H$-$T$ phase-diagram in the inset of Fig.~\ref{fig:ac_chi}(d). While the suppression of $T_f$ by fields is similar to the cooperative freezing behavior in spin-glasses, the increase of $T_f$ at low $H$ cannot be understood by spin-glass physics.  An increase of $T_f$ with field has been observed in the spin-ice systems Dy$_2$Ti$_2$O$_7$ \cite{Snyder2004} and Ho$_2$Ti$_2$O$_7$ \cite{Ehlers2004}. The complex field dependence of $T_f$ reflects a complex spin dynamics in Tb$_2$Hf$_2$O$_7$.

\section{\label{Sec:ChiMH}  \lowercase{dc} Magnetic Susceptibility and Magnetization}

\begin{figure}
\includegraphics[width=\columnwidth, keepaspectratio]{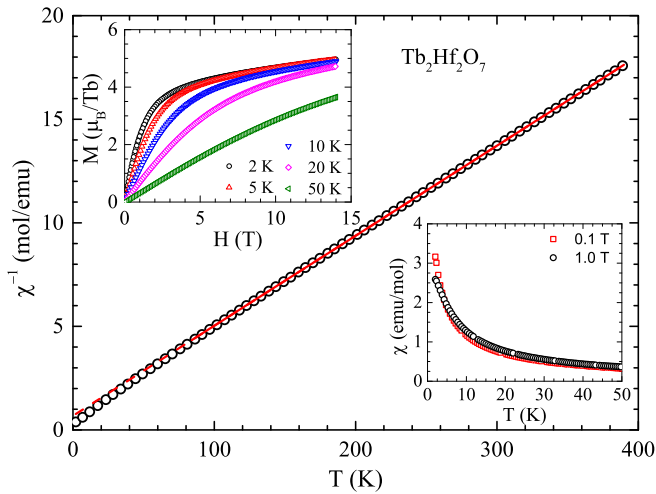}
\caption{Zero-field-cooled magnetic susceptibility $\chi$ of polycrystalline Tb$_2$Hf$_2$O$_7$ plotted as inverse magnetic susceptibility $\chi^{-1}$ as a function of temperature $T$ for $2~{\rm K} \leq T \leq 390$~K measured in magnetic field $H= 1.0$~T\@. The solid red line is a Curie-Weiss fit for $50~{\rm K} \leq T \leq 390$~K and the dashed line is the extrapolation of the fit towards 0~K\@. Lower inset: Low-$T$ $\chi(T)$ data measured in $H= 0.1$~T and 1.0~T\@. Upper inset: Isothermal magnetization $M(H)$ for $0 \leq H \leq 14$~T measured at the indicated temperatures.} 
\label{fig:MT}
\end{figure}

The zero-field-cooled (ZFC) dc magnetic susceptibility $\chi(T)$ of Tb$_2$Hf$_2$O$_7$ is shown in Fig.~\ref{fig:MT} for $2~{\rm K} \leq T \leq 390$~K\@. Consistent with $\chi_{\rm ac}(T)$, no anomaly related to a magnetic phase transition is seen in $\chi(T)$ data, though the magnitude of $\chi$ at low-$T$ is rather large (see the lower inset of Fig.~\ref{fig:MT}) as commonly seen in the $R_2B_2$O$_7$ pyrochlores. No thermal hysteresis was observed between the ZFC and field-cooled (FC) $\chi(T)$ data (not shown) above 2~K\@. The $\chi(T)$ data follow a Curie-Weiss behavior, $\chi(T) =  C/(T-\theta_{\rm p})$. 
A fit of the ZFC $\chi^{-1}(T)$ data between 50 and 390~K, measured in 1.0~T, shown by the solid red line in Fig.~\ref{fig:MT}, yields $C = 11.47(3)$~emu\,K/mol\,Tb and $\theta_{\rm p} = -14.6(5)$~K\@. The effective moment obtained from $C$, $\mu_{\rm eff} \approx 9.58(2)\, \mu_{\rm B}$/Tb, is very close to the expected value of $9.72 \, \mu_{\rm B}$ for free Tb$^{3+}$ ions.  The negative value of $\theta_{\rm p}$ indicates an antiferromagnetic interaction in Tb$_2$Hf$_2$O$_7$. The $\chi^{-1}(T)$ data  when fitted in the $T$ range of 15 to 30~K yields $C = 10.03(2)$~emu\,K/mol\,Tb and $\theta_{\rm p} = -6.10(4)$~K\@, thus an effective moment of $8.96~\mu_{\rm B}$ is obtained for the low temperature ground state.

The isothermal magnetization $M(H)$ collected at five selected temperatures between 2 and 50~K is shown in the upper inset of Fig.~\ref{fig:MT}. At low temperatures, $M$ increases rapidly at low $H$ and weakly at higher $H$. At 2~K and 14~T the magnetization attains a value of $M \approx 4.9\, \mu_{\rm B}$/Tb which is only about 54\% of the theoretical saturation magnetization of $9\,\mu_{\rm B}$/Tb for free Tb$^{3+}$ ions. Such a low value of $M$ suggests the presence of strong single-ion anisotropy. Further it is seen that the $M(H)$ shows slight nonlinearity even at 50~K which can be attributed to the combined effect of crystal field and saturation tendency. We also notice that even at 2~K the $M$ does not fully saturate up to 14~T, instead shows a weak increase with increasing $H$. This could be the effect of CEF and/or presence of disorder. 

We would like to point out that the $M(H)$ data could not be described by the effective spin-half model of local $\langle 111 \rangle$ Ising anisotropy observed in Ising pyrochlores such as in Dy$_2$Ti$_2$O$_7$ \cite{Bramwell2000}, Nd$_2$Zr$_2$O$_7$ \cite{Xu2015}  and Nd$_2$Hf$_2$O$_7$ \cite{Anand2015}. It appears that the presence of anion disorder modifies the crystal field anisotropy leading to the departure from the expected Ising-type anisotropy in Tb$_2$Hf$_2$O$_7$. 

\begin{figure}
\includegraphics[width=\columnwidth, keepaspectratio]{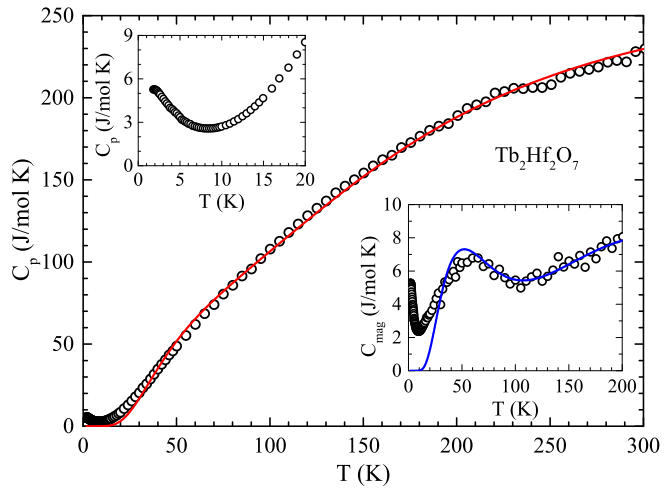}
\caption{\label{fig:HC} Heat capacity $C_{\rm p}$ of polycrystalline Tb$_2$Hf$_2$O$_7$ as a function of temperature $T$ between 1.8 and 300~K, measured in zero field. The solid red curve is the fit by Debye + Einstein models of lattice heat capacity plus crystal field contribution. Upper inset: Expanded view of \mbox{low-$T$} $C_{\rm p}(T)$ data between 1.8 and 20~K\@. Lower inset: Magnetic contribution to heat capacity $C_{\rm mag}(T)$ of Tb$_2$Hf$_2$O$_7$. The solid blue curve represents the crystal-field contribution to heat capacity.} 
\end{figure}

\section{\label{Sec:HC}Heat Capacity}

The heat capacity $C_{\rm p}(T)$ of Tb$_2$Hf$_2$O$_7$ is shown in Fig.~\ref{fig:HC} for 1.8~K~$\leq T \leq$~300~K\@. The $C_{\rm p}(T)$ data show no anomaly related to a phase transition down to 1.8~K. However, as visualized in the upper inset, the $C_{\rm p}(T)$ data show an upturn below about 8 K that might indicate the precursor of magnetic order.  Further it is seen that the $C_{\rm p}(T)$ tends to flatten below 2~K, possibly indicating the occurence of a peak below 2~K\@. 

The magnetic contribution to the heat capacity $C_{\rm mag}(T)$, which was estimated by subtracting off the lattice contribution, is shown in the lower inset of Fig.~\ref{fig:HC}. For the lattice contribution we used the heat capacity of nonmagnetic La$_2$Hf$_2$O$_7$ \cite{Anand2015} after correcting for the small difference in formula masses and unit cell volumes of La$_2$Hf$_2$O$_7$ and  Tb$_2$Hf$_2$O$_7$. The $C_{\rm mag}(T)$ exhibits a broad Schottky-type peak near 55~K\@. The $C_{\rm mag}(T)$ data below 200~K is reasonably described by a three-level CEF scheme. The analysis of $C_{\rm mag}(T)$ data  suggests the ground state to be a doublet with a first excited doublet at around 125~K and a quasi-quartet state at around 710~K\@. The CEF contribution to heat capacity $C_{\rm CEF}(T)$ is shown by the solid blue curve in the lower inset of Fig.~\ref{fig:HC}. A nice agreement is seen between $C_{\rm mag}(T)$ and $C_{\rm CEF}(T)$ above 20~K where the interactions between Tb$^{3+}$ ions are insignificant and the Schottky type feature is well accounted for. 

 The inelastic neutron scattering data of Tb$_2$Hf$_2$O$_7$ do not show discrete CEF excitations, instead broad inelastic signals (mostly between 5 and 25 meV) are observed \cite{Sibille2017}, which can be attributed to a distribution of crystal field environments around Tb$^{3+}$. Sibille {\it et al}.\ \cite{Sibille2017} suggested an energy gap of at least 4.3 meV (50 K) for the first excited state. The raw inelastic neutron scattering data show a rather broad CEF excitation between 5 and 10 meV (associated with the first excited state), which is somewhat smaller than our estimate of 11 meV (125 K) from the heat capacity.

The high temperature $C_{\rm p}(T)$ data show that the $C_{\rm p}$ does not reach the Dulong-Petit value of $C_{\rm V} = 3 nR = 33 R \approx 274.4$ J/mol\,K by 300~K ($C_{\rm p} \sim 225$ J/mol\,K). This is consistent with the large value of Debye temperature $\Theta_{\rm D}$ observed in the $R_2B_2$O$_7$ pyrochlore family. A fit of the $C_{\rm p}(T)$ data by a combination of the Debye and Einstein models of lattice heat capacity added to the fixed crystal field contribution $C_{\rm CEF}(T)$ (according to the CEF level scheme discussed above) \cite{Anand2015a} shown by the solid red curve in Fig.~\ref{fig:HC} gives $\Theta_{\rm D} = 787(6)$~K and Einstein temperature $\Theta_{\rm E} = 163(3)$~K with 66\% weight to Debye term and 34\% to Einstein term.  The value of $\Theta_{\rm D}$ obtained compares with that of the other pyrohafnates La$_2$Hf$_2$O$_7$ ($\Theta_{\rm D} = 792(5)$~K) \cite{Anand2015}, Nd$_2$Hf$_2$O$_7$ ($\Theta_{\rm D} = 785(6)$~K) \cite{Anand2015} and Pr$_2$Hf$_2$O$_7$ ($\Theta_{\rm D} = 790(7)$~K) \cite{Anand2016}. 

\section{\label{Sec:ND}  Neutron Powder Diffraction: Diffuse Scattering}

\begin{figure}
\includegraphics[width=\columnwidth, keepaspectratio]{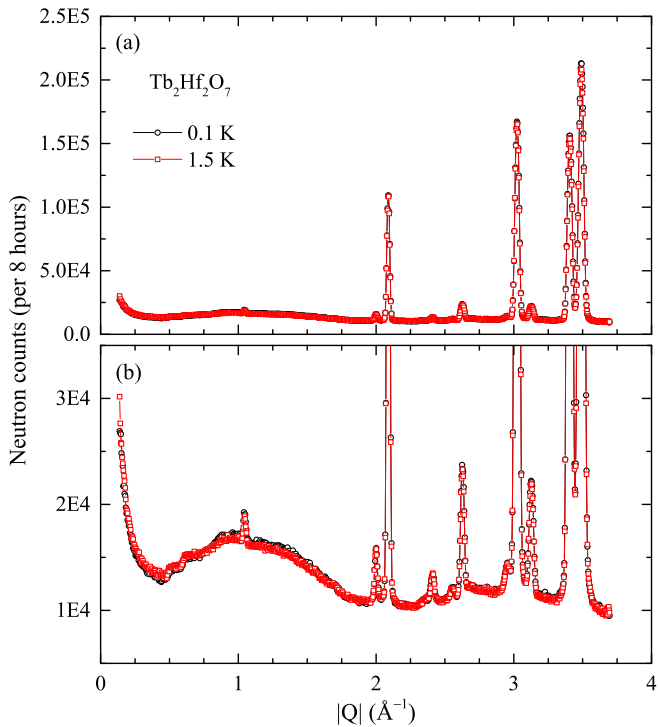}
\caption{\label{fig:ND} (a) Neutron powder diffraction patterns of Tb$_2$Hf$_2$O$_7$ collected at 0.1 and 1.5~K. (b) Expanded $y$-scale plot showing the broad shoulder-like feature at low $|Q|$ due to magnetic diffuse scattering.} 
\end{figure}

\begin{figure}
\includegraphics[width=\columnwidth, keepaspectratio]{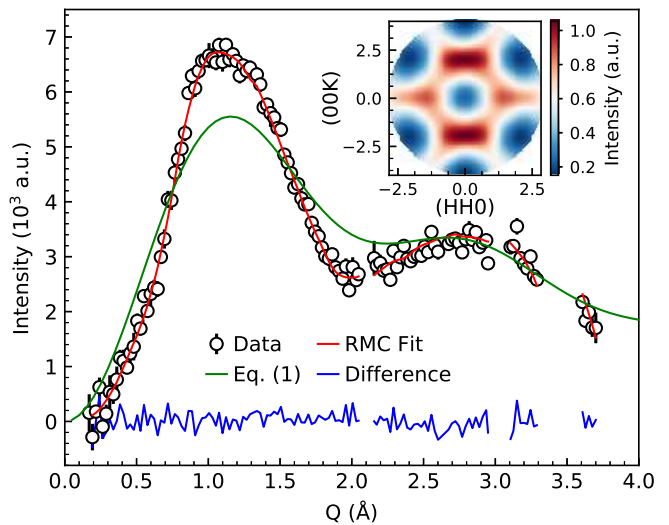}
\caption{\label{fig:ND_RMC} (Color online) :  Magnetic diffuse neutron  scattering pattern of Tb$_2$Hf$_2$O$_7$ at 100 mK obtained by subtracting the 160~K data from the data at 100~mK, calculated powder-averaged scattering pattern due to nearest-neighbor isotropic AFM spin correlations (green line), and reverse Monte Carlo (RMC) fit (red line) and the difference (blue line). Inset: The reconstructed diffraction pattern in the (hhl) reciprocal plane based on the fitted spin configurations.} 
\end{figure}

Neutron powder diffraction patterns of Tb$_2$Hf$_2$O$_7$ collected at 0.1 and 1.5~K are shown in Fig.~\ref{fig:ND}(a). No magnetic Bragg peaks are observed in ND data down to 0.1~K, however a broad shoulder-like feature is seen at lower $Q$ which is more clearly seen in the expanded plot shown in Fig.~\ref{fig:ND}(b). This broad shoulder arises due to magnetic diffuse scattering  as a result of the development of short-range magnetic correlations at low temperatures. A similar feature in ND data has also been observed for the spin-liquid Tb$_2$Ti$_2$O$_7$ \cite{Gardner1999,Gardner2001}. In order to better resolve the magnetic diffuse scattering we subtracted the ND data taken at 160~K from those taken at 0.1~K\@. The  magnetic intensity at 0.1~K is shown in Fig.~\ref{fig:ND_RMC} (the featureless paramagnetic scattering was added according to \cite{Paddison2015} after subtraction). The sharp decrease of diffuse scattering intensity when $Q$ is approaching zero indicates dominant antiferromagnetic correlations, consistent with the negative Weiss temperature.

The broad peak near $Q\sim1.2$~\AA$^{-1}$ (Fig.~\ref{fig:ND_RMC}) corresponds to the nearest-neighbor spin correlations with a distance 3.7~\AA. The large peak width indicates that the spin correlations exist only over a very short range. The powder-averaged scattering due to the nearest-neighbor isotropic spin correlations could be described by \cite{Bertaut1967}:
\begin{equation}
I(Q) \sim f (Q) ^2  \frac{\sin(Qr_{ij})}{Qr_{ij}},
\end{equation}
where $f(Q) ^2$ is the magnetic form factor of Tb$^{3+}$ and  $r_{ij}\approx3.7$\AA\ is the nearest-neighbor distance. 
As shown in Fig.~\ref{fig:ND_RMC}, the model qualitatively accounts for the modulation of the intensity which suggests that the spin correlations only extend over nearest neighbors.

In order to gain further insight on the short-range spin correlations and spin anisotropy in Tb$_2$Hf$_2$O$_7$, the magnetic diffuse  scattering pattern was analyzed with the reverse Monte Carlo (RMC) method. The RMC algorithm describes the experimental powder magnetic diffuse scattering data by a large configuration of classical spin vectors with certain anisotropy. During the refinement, the orientations of the spins are refined in order to describe best the experimental data. The program SPINVERT \cite{Paddison2013} is used which has been successfully applied to several frustrated magnetic systems and reveals many before-unnoticed novel types of spin correlations \cite{Paddison2013, Paddison2012, Nilsen2015, Paddison2014}.  In our calculations, a $6\times6\times6$  supercell is generated (in total 3456 spins) with randomly oriented magnetic moments (isotropic spins) assigned to each magnetic Tb$^{3+}$ site. In total 500 moves per spin are considered for a refinement and ten individual fits have been performed to ensure the robustness of the results. 

The fitted magnetic diffuse scattering pattern with isotropic spins is shown in Fig.~\ref{fig:ND_RMC}, which agrees well with the experimental data. The resulting spin configurations were used to reconstruct the $Q$-dependence of the diffuse scattering in the ($hhl$) reciprocal plane (inset of Fig.~\ref{fig:ND_RMC}) by using the SPINDIFF program which is an extension to the SPINVERT program.  The calculated pattern is quite similar to the experimentally observed pattern obtained from the single crystal diffuse neutron scattering of Tb$_2$Hf$_2$O$_7$ reported by Sibille {\it et al}. \cite{Sibille2017}. The fitted spin configurations reveal that the spins are generally antiferromagnetically correlated, which is consistent with the negative Weiss temperature. 

We would like to mention that we also performed refinements with spins of local $\langle 111 \rangle$  Ising-anisotropy as well as easy-plane anisotropy. The Ising anisotropy case did not yield an acceptable fit. Considering easy-plane anisotropy is also found to give a fit quality similar to the one with isotropic spins, however, the calculated single crystal scattering pattern based on the refined spin configurations differs significantly from the observed pattern in Ref.~\cite{Sibille2017}. This leaves the refinement with isotropic spins as the only acceptable solution which reflects the non-Ising character of anisotropy of Tb$^{3+}$ in Tb$_2$Hf$_2$O$_7$, possibly because of the mixing of the CEF ground state with excited states by exchange interactions and the structural disorder.

\begin{figure}
\includegraphics[width=\columnwidth]{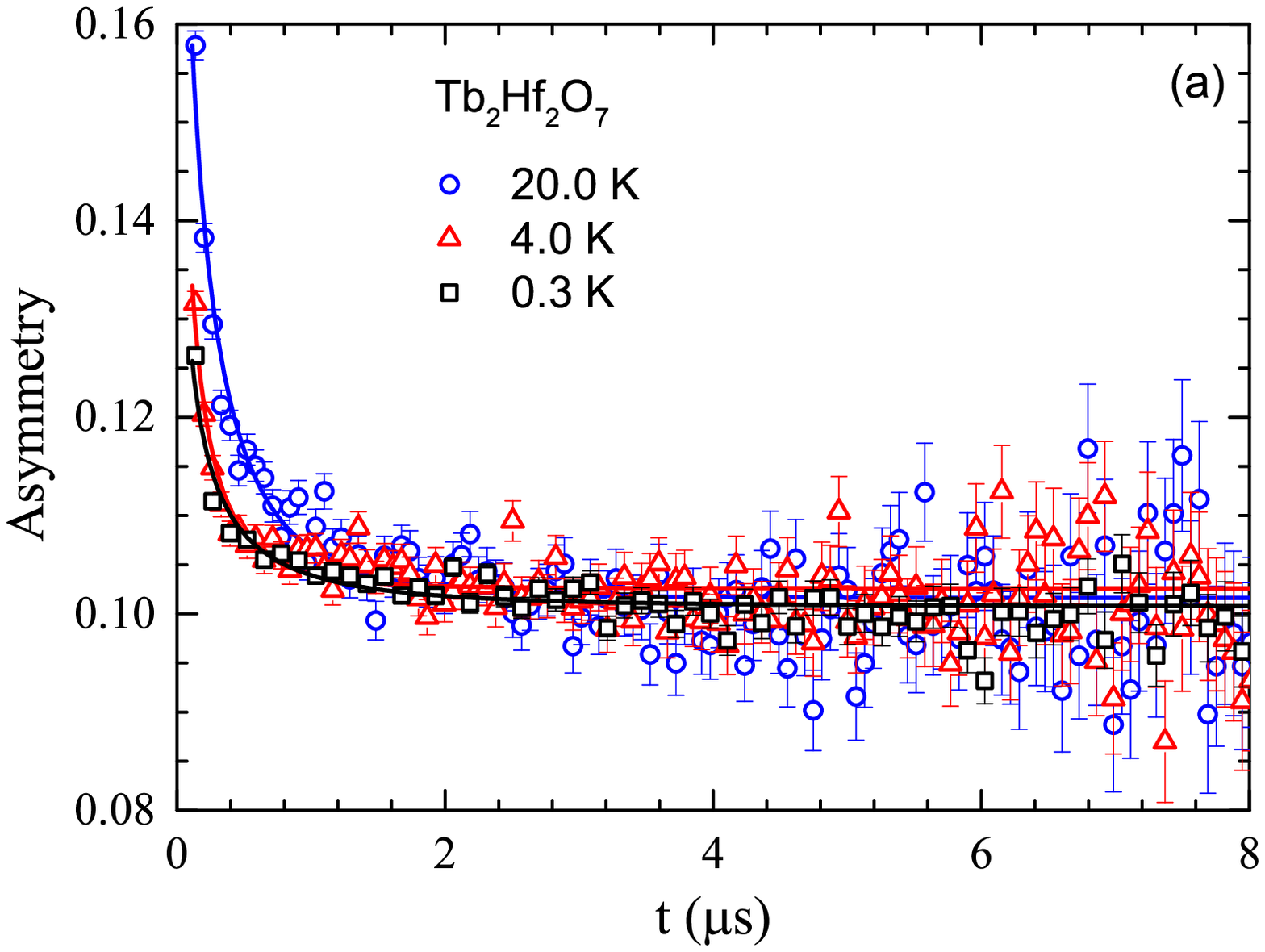}
\includegraphics[width=\columnwidth]{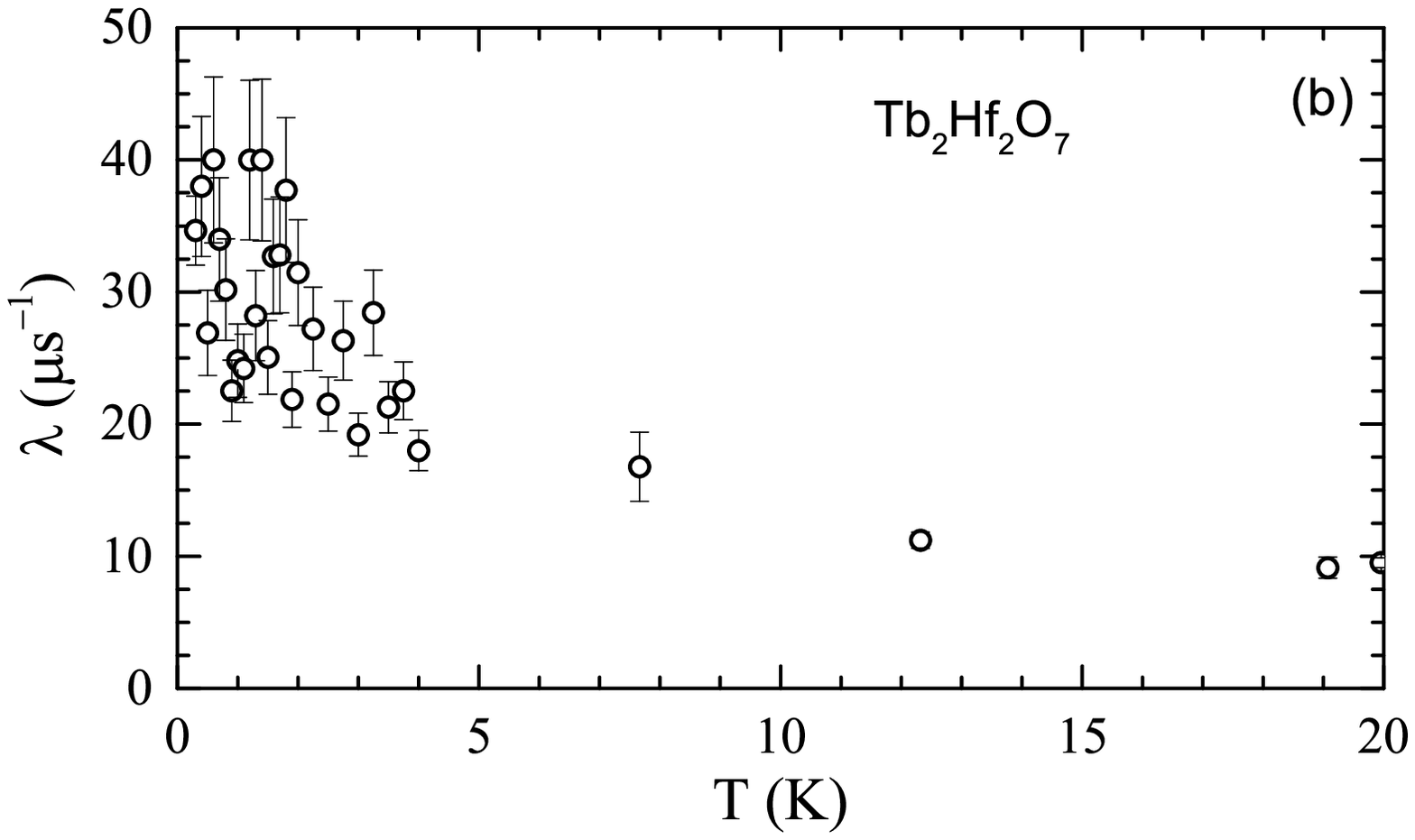}
\includegraphics[width=\columnwidth]{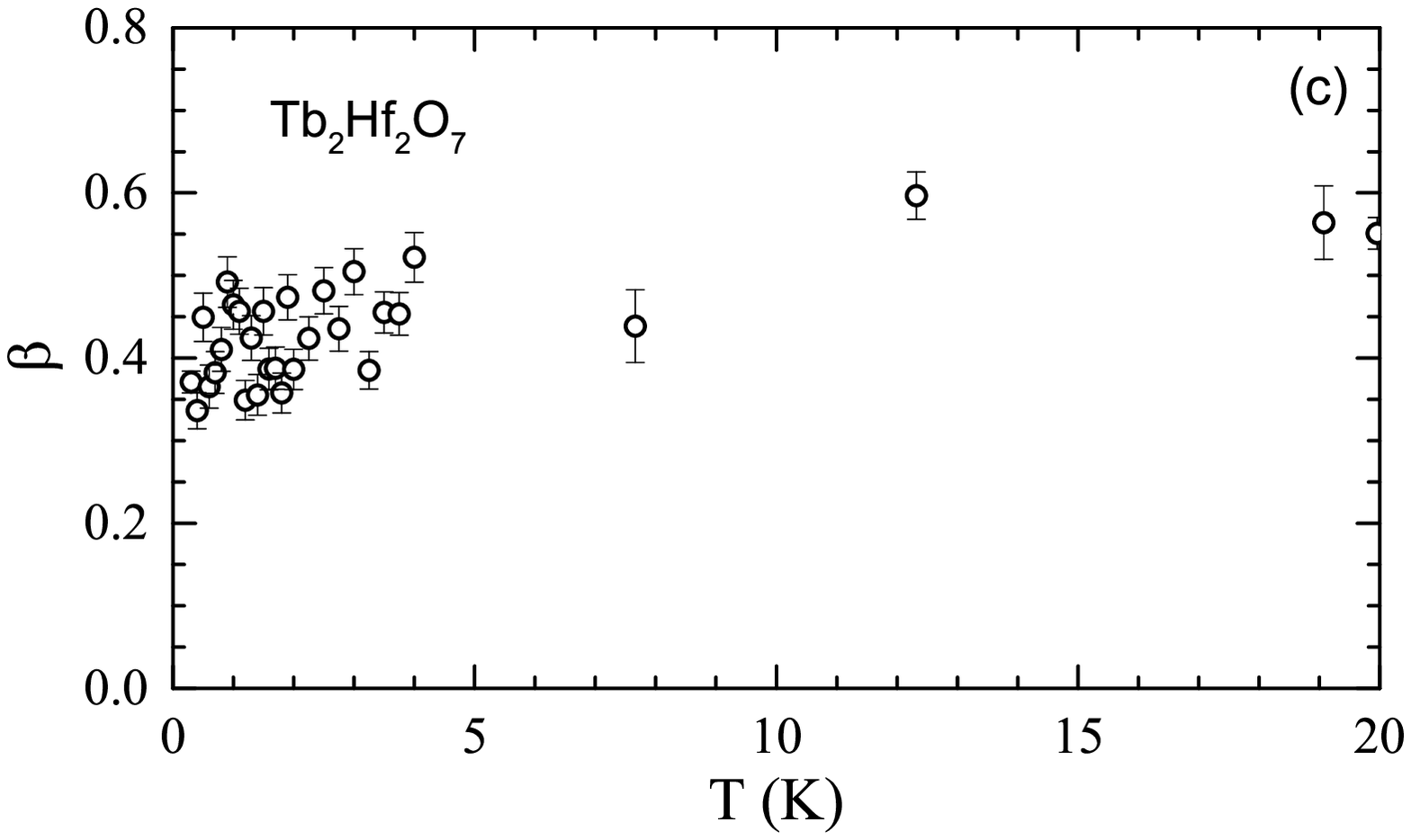}
\caption{\label{fig:MuSR1} (a) Zero field (ZF) muon spin asymmetry function $G_z$ versus time $t$ spectra of Tb$_2$Hf$_2$O$_7$ at few representative temperatures. Solid curves are the fits to the $\mu$SR data by a stretched exponential relaxation function, $G_z(t) = A_{0} \exp[({-\lambda t}) ^\beta]+ A_{\rm BG}$. (b) Temperature $T$ dependence of depolarization rate $\lambda$, and (c) $T$ dependence of exponent  $\beta$, obtained from the analysis of the ZF $\mu$SR data at $0.3\leq T \leq 20$~K\@.}
\end{figure}

\begin{figure}
\includegraphics[width=\columnwidth]{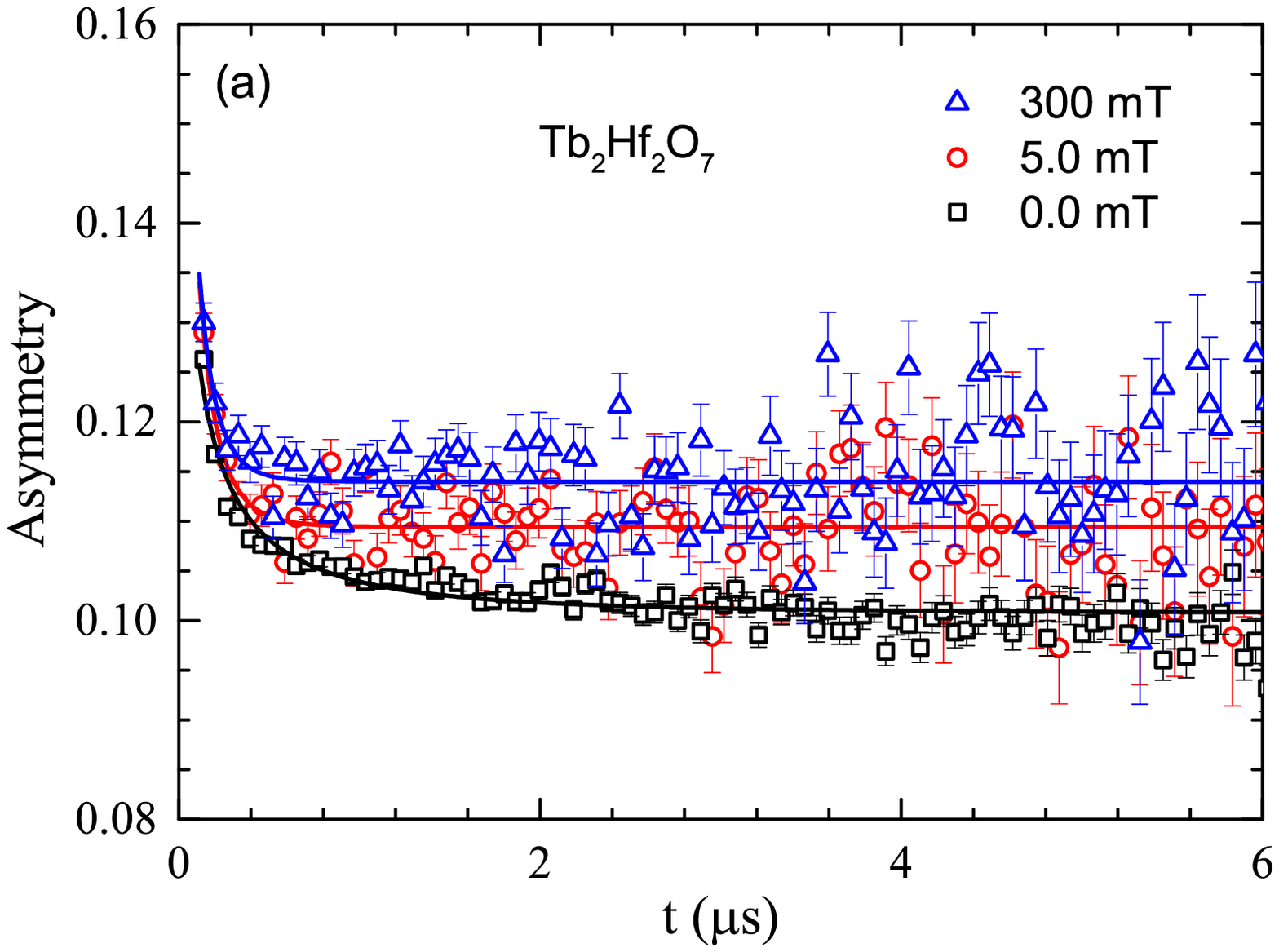}
\includegraphics[width=\columnwidth]{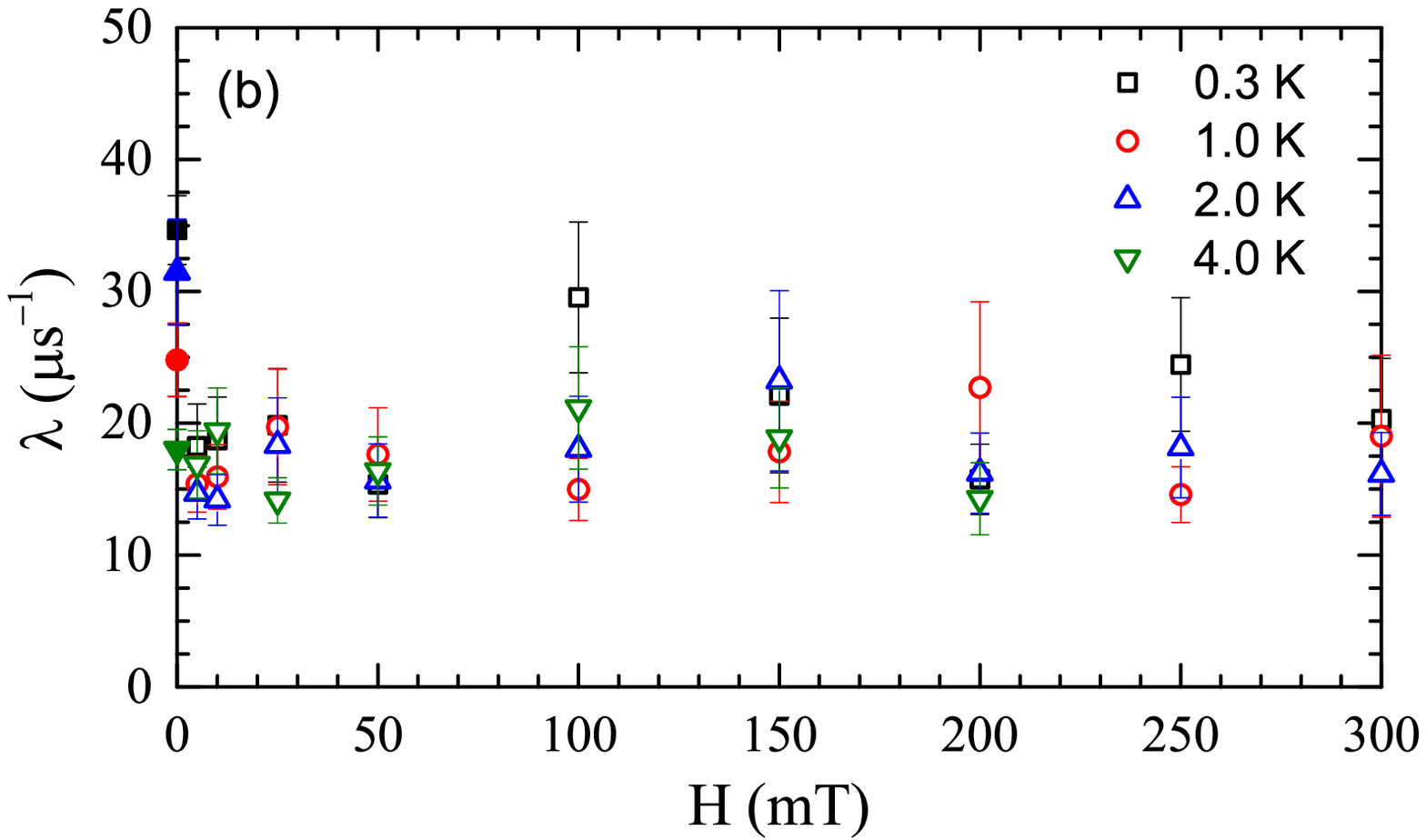}
\includegraphics[width=\columnwidth]{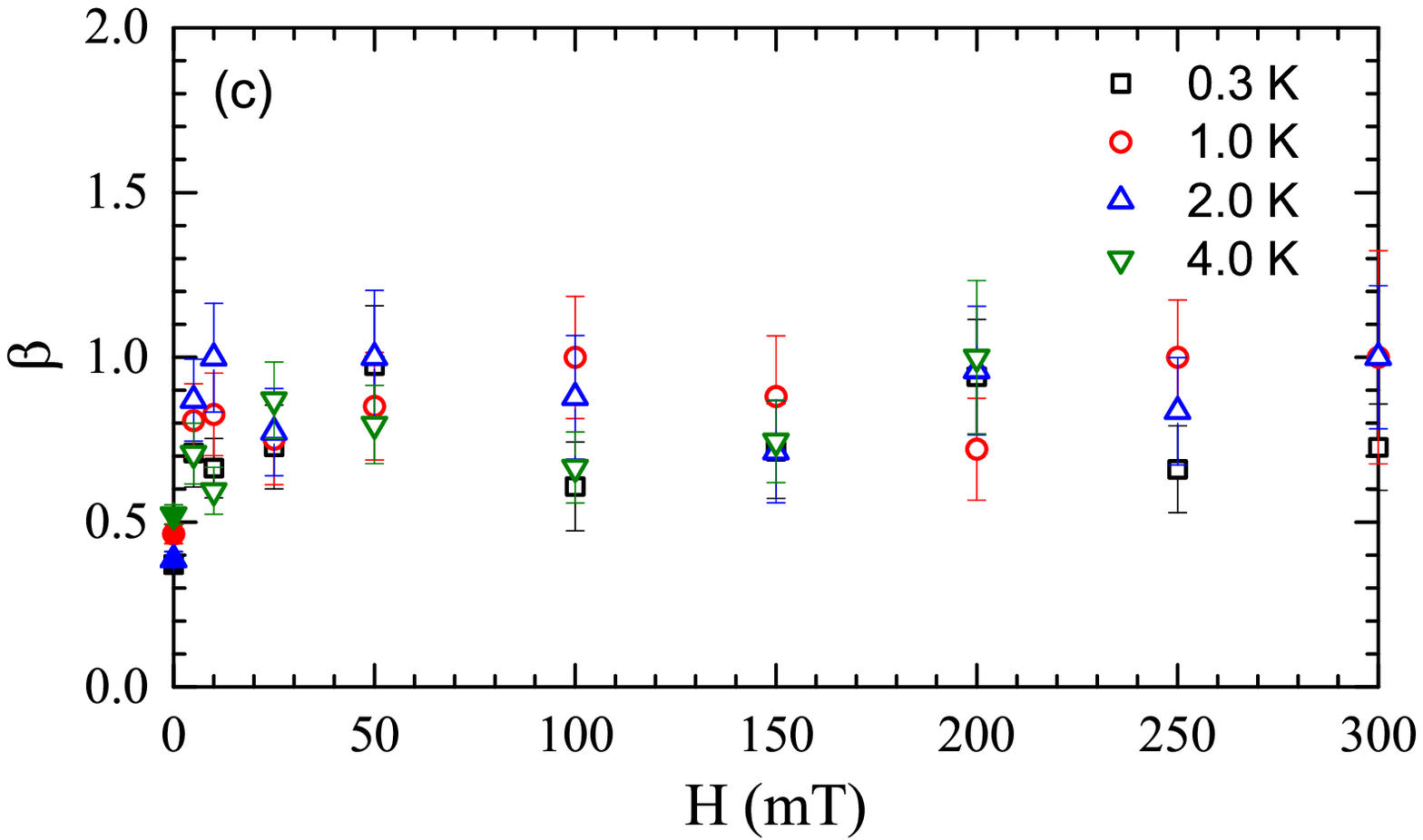}
\caption{\label{fig:MuSR2} (a) Longitudinal-field (LF) muon spin asymmetry function $G_z$ versus time $t$ spectra of Tb$_2$Hf$_2$O$_7$ at 0.3~K for a few representative fields. Solid curves are the fits to the $\mu$SR data by a stretched exponential relaxation function, $G_z(t) = A_{0} \exp[({-\lambda t}) ^\beta]+ A_{\rm BG}$. (b) Magnetic field $H$ dependence of the depolarization rate $\lambda$, and (c) $H$ dependence of the exponent  $\beta$, obtained from the analysis of the LF $\mu$SR data. The data points at $H=0$ (solid symbols) in (b) and (c) correspond to the parameters $\lambda$ and $\beta$ for zero field $\mu$SR data inFig.~\ref{fig:MuSR1}.}
\end{figure}

\section{\label{Sec:muSR} Muon Spin Relaxation}

In order to obtain further insight to the spin dynamics of Tb$_2$Hf$_2$O$_7$, we also carried out  $\mu$SR measurements.  The  $\mu$SR data were reduced and analyzed using the program Mantid \cite{Arnold2014}. The $\mu$SR spectra of Tb$_2$Hf$_2$O$_7$ collected in zero field (ZF) are shown in Fig.~\ref{fig:MuSR1} for a few representative temperatures between 0.3 and 20~K\@. The $\mu$SR spectra do not show any signature of long-range magnetic order (no oscillations related to muon spin precession)  or a static ground state (no 2/3 loss in asymmetry compared to its high-$T$ value). Thus, consistent with the above neutron data, the muons also do not see a magnetic ordering down to 0.3~K.  The $\mu$SR spectra are well described by a stretched exponential relaxation function, $G_z(t) = A_{0} \exp[({-\lambda t}) ^\beta]+ A_{\rm BG}$, where $A_{0}$ is the initial asymmetry, $\lambda$ is the depolarization rate and $\beta$ is the exponent. In the fast fluctations limit, $\lambda$ is related to the spin-fluctuation rate $\nu$ which in Redfield formalism is given by \cite{Yaouanc2011}, 
\begin{equation}
 \lambda = \frac{2\gamma_\mu^2  H_{\mu}^2  \nu}{\nu^2+\gamma_\mu^2 H_{\rm LF}^2 },
\label{eq:Lambda}
\end{equation}
where $\gamma_\mu$ is the muon gyromagnetic ratio, $H_{\mu}$ is the magnitude of the fluctuating field at the muons site, and $H_{\rm LF}$ is the applied longitudinal field. For a dynamic relaxation one expects $\beta=1$ (exponential relaxation function) whereas for the static or quasi-static case one would expect $\beta=2$. When there is a distribution of dynamic relaxation channels, $\beta$ is less than one.  

The fits of the representative $\mu$SR spectra are shown by the solid curves of respective colors in Fig.~\ref{fig:MuSR1}(a). The parameters $\lambda$ and $\beta$ obtained from the fits are shown in Figs.~\ref{fig:MuSR1}(b) and \ref{fig:MuSR1}(c), respectively. It is seen that $\lambda$ tends to increase with decreasing temperature. According to Eq.~(\ref{eq:Lambda}), for zero field $\lambda \sim 1/\nu$, therefore an increase in $\lambda$ implies a decrease in the spin-fluctuation rate and, hence, a slowing down of spin dynamics. The value of $\beta$ is much lower than 1 and suggests a distribution of relaxation channels. Thus, from the ZF $\mu$SR data we conclude the existence of slow spin dynamics and persistent spin fluctuations. Our findings are similar to those of Sibille {\it et al}.\ \cite{Sibille2017} who also reported a slowing down of spin dynamics and coexisting spin fluctuations in the macroscopically frozen state of Tb$_2$Hf$_2$O$_7$ through their ZF $\mu$SR study.

Figure~\ref{fig:MuSR2}(a) shows the longitudinal-field (LF) $\mu$SR asymmetry spectra of Tb$_2$Hf$_2$O$_7$ for  a few representative fields between 5 and 300~mT collected at 0.3~K\@. For comparison the ZF $\mu$SR data at 0.3~K are also shown. From the raw spectra [Fig.~\ref{fig:MuSR2}(a)] we see that there is almost no change in the initial asymmetry for fields up to 300~mT. The LF $\mu$SR spectra are also well described by the stretched exponential relaxation function. The fits are shown by the solid curves in Fig.~\ref{fig:MuSR2}(a). The fit parameters for the LF data collected at 0.3, 1.0, 2.0, and 4.0 are shown in Figs.~\ref{fig:MuSR2}(b) and \ref{fig:MuSR2}(c). A comparison of the fit parameters for ZF spectra in Fig.~\ref{fig:MuSR1} and LF spectra in Fig.~\ref{fig:MuSR2} shows that with the application of field the values of both $\lambda$ and $\beta$ change. While $\lambda$ decreases, $\beta$ increases which is noticed even at 5.0~mT field. Further it is seen that both $\lambda$ and $\beta$ show very weak $H$ dependence (almost $H$-independent within the error bar)  over 5~mT to 300~mT. Thus, the $H$ dependent behavior of the muon relaxation does not follow Eq.~(\ref{eq:Lambda}) according to which one would expect a decrease in $\lambda$ with increasing $H_{\rm LF}$. This suggests that the spin dynamics in Tb$_2$Hf$_2$O$_7$ is more complicated than the one described by Eq.~(\ref{eq:Lambda}). We further notice that fields up to 300~mT are not sufficient to restore the initial polarization. This suggests a dynamic nature of the muon relaxation in Tb$_2$Hf$_2$O$_7$. Much higher fields have been suggested to restore the initial polarization when the muon relaxation is dynamic in nature \cite{Bert2006}.

\section{\label{Conclusion} Conclusions}

We have investigated the physical properties of Tb$_2$Hf$_2$O$_7$ using $\chi_{\rm ac}(T)$, $\chi(T)$, $M(H)$, $C_{\rm p}(T)$, $\mu$SR, and neutron powder diffraction measurements. The $\chi(T)$ and $C_{\rm p}(T)$ show no anomaly related to a magnetic phase transition down to 1.8~K\@. The negative $\theta_{\rm p}$ obtained from the analysis of $\chi(T)$ indicates a dominant antiferromagnetic interaction. The $M(H)$ data reflect strong single-ion anisotropy. The $C_{\rm p}(T)$ shows an upturn below 8~K\@. Clear evidence for a slowing down of spin dynamics is provided by $\chi_{\rm ac}(T)$ data. The frequency- and field-dependent $\chi_{\rm ac}(T)$ data reveal complex spin dynamics in Tb$_2$Hf$_2$O$_7$. The $\mu$SR data follow a stretched exponential behavior reflecting a slow spin dynamics and persistent spin fluctuations down to 0.3~K with a distribution of relaxation channels\@. The $\chi_{\rm ac}(T)$ and $\mu$SR data together provide evidence for coexisting persistent dynamic spin fluctuations inside the macroscopically spin-frozen state. The powder neutron diffraction show magnetic diffuse scattering which has been analyzed by the RMC method and a dominant antiferromagnetic correlation is inferred. The RMC analysis also shows approximately isotropic spins rather than Ising anisotropy. Altogether our investigations reveal a dynamical ground state in Tb$_2$Hf$_2$O$_7$.

The evidence for the existence of a dynamical ground state found by us, is consistent with the recently proposed Coulomb spin-liquid behavior by Sibille {\it et al}.\ \cite{Sibille2017}. Sibille {\it et al}.\ \cite{Sibille2017} as well observe a frequency-dependent $\chi_{\rm ac}(T)$ and a non-Arrhenius behavior for spin freezing in Tb$_2$Hf$_2$O$_7$. They suggested a spin-glass type freezing, which, however, is not supported by our $\chi_{\rm ac}(T)$ data collected with applied dc magnetic field. Sibille {\it et al}.\ \cite{Sibille2017} also find evidence for a dynamic nature of magnetism in Tb$_2$Hf$_2$O$_7$ in their $\mu$SR study. They suggested a macroscopically frozen state to coexist with fast fluctuations which we as well infer from our $\mu$SR results. Sibille {\it et al}.\ \cite{Sibille2017}  further observe magnetic diffuse scattering as a result of the development of short-range spin-spin correlations. This is again consistent with our observation of diffuse scattering. In addition, Sibille {\it et al}.\ \cite{Sibille2017}  probe spin correlations using low-energy inelastic neutron scattering which provides evidence for a Coulomb phase, characterized by power-law correlations.

Our investigations suggest the ground state properties of Tb$_2$Hf$_2$O$_7$ to have some similarities with those of the spin-liquid candidate Tb$_2$Ti$_2$O$_7$, though the pyrochlore structure of Tb$_2$Hf$_2$O$_7$ is found to have disorder. They both have dominating antiferromagnetic interaction, show non-Arrhenius behavior in ac susceptibility measurement, diffuse magnetic scattering in neutron diffraction measurement, and stretched exponential behavior in muon spin relaxation measurement. On the other hand,  magnetic heat capacity as well as inelastic neutron scattering data of Tb$_2$Hf$_2$O$_7$ \cite{Sibille2017} suggest a first excited state much higher than that in Tb$_2$Ti$_2$O$_7$ for which the first excited CEF level at around 1.4~meV has been proposed to renormalize the effective low-energy spin Hamiltonian leading to a quantum spin-ice state in Tb$_2$Ti$_2$O$_7$ \cite{Molavian2007,Molavian2012}. It appears that the presence of disorder plays an important role in the dynamical behavior of Tb$_2$Hf$_2$O$_7$.  As such Tb$_2$Hf$_2$O$_7$ seems to be a potential compound for further investigations to understand the role of disorder on the spin dynamics in these pyrochlores. 

\acknowledgements
We acknowledge the Helmholtz Gemeinschaft for funding via the Helmholtz Virtual Institute (Project No. VH-VI-521) and DFG through SFB 1143. We also acknowledge support by HLD at HZDR, member of the European Magnetic Field Laboratory (EMFL). E.C. acknowledges support from the Danish Research Council for Science and Nature through DANSCATT.

\end{document}